\documentstyle[epsf,aps]{revtex}

\def\be{\begin{equation}}
\def\ee{\end{equation}}
\def\bea{\begin{eqnarray}}
\def\eea{\end{eqnarray}}

\begin{document}

\twocolumn[\hsize\textwidth\columnwidth\hsize\csname
@twocolumnfalse\endcsname

\title{Low energy effective theory for two branes system \\
-- Covariant curvature formulation --}
\author{Tetsuya Shiromizu$^{(1,2)}$ and Kazuya Koyama$^{(3)}$}

\address{$^{(1)}$ Department of Physics, Tokyo Institute of Technology, Tokyo 152-8551, Japan}

\address{$^{(2)}$ Advanced Research Institute for Science and Engineering,
Waseda University, Tokyo 169-8555, Japan
}

\address {$^{(3)}$ Department of Physics, The University of Tokyo, Tokyo 113-0033, Japan}

\date{\today}

\maketitle

\begin{abstract}
We derive the low energy effective theory for two branes system solving the bulk geometry 
formally in the covariant curvature formalism developed by 
Shiromizu, Maeda and Sasaki. As expected, the effective 
theory looks like a Einstein-scalar system. Using this theory we can discuss the cosmology and 
non-linear gravity at low energy scales.
\end{abstract}
\vskip2pc]


\vskip1cm


\section{Introduction}

The recent progress in superstring theory provides us the new picture of the universe, 
so called brane world, where our universe is like domain wall in higher dimensional spacetimes. 
The matter is confined on the 3-brane. The simplest model was proposed by Randall and 
Sundrum\cite{RSI,RSII}. In their model the bulk spacetimes is 5-dimensional anti-deSitter 
spacetimes and the brane is four dimensional Minkowski spacetimes. Their first model(RS1) 
gives us the geometrical solution to the gauge hierarchy problem. The RS1 model consists of 
two brane, the positive and negative tension branes. For the gauge hierarchy problem, 
it is supposed that the visible brane where we are is the negative tension one. The linealised theory 
has been carefully investigated in \cite{Tama,Rubakov}(See Ref. \cite{Csaba,Gen,Kazuya} for 
the cosmological cases. See Ref \cite{Other} for another issues.). As a result, it is turned out 
that the gravity on the brane looks like the scalar-tensor gravity. However, we have no 
successful analysis on the non-linear aspect of the gravity except for the second order 
perturbation\cite{Kudoh2}. Recently there are impressive 
progress\cite{Toby,Kanno,Kanno2} on this issue. In Ref. \cite{Kanno}, the effective 
equation is derived at the low energy scale. 

In this paper we re-derive the effective theory for two brane systems which was obtained by 
Kanno and Soda\cite{Kanno}(See also \cite{Toby}) in the metric based approach. On the other hand, 
our approach is based on the covariant 
curvature formalism\cite{Tess,Roy}. As seen later,  
our derivation is much simpler than the metric based approach and straightforward. 
The covariant curvature formalism gives us a gravitational equation on the branes. 
For RS2 models which consists of the single brane, this is powerful approach to look 
at the full view of the brane world. Indeed, it was easy to see that the Newton gravity is recovered at the 
low energy. The approach, however, is not regarded as so useful for RS1 models. 
We shall show this is not true. This is the main purpose in this paper. For simplicity, 
we will not address the stabilisation issue of two branes(radion stabilisation problems)
\cite{Wise,Tama2,Csaba2}. 

The rest of this paper is organised as follows. In the section 2, we summarise the 
covariant curvature formalism. In the section 3, we formally solve the bulk 
perturbatively up to the 1st order. The infinitesimal and dimensionless parameter is the ratio of the 
bulk to brane curvature radii. After then we derive the effective theory at the 
1st order. The equation includes the non-linear part of the induced gravity on the 
branes. In the section 4, we summarise the present work.

\section{Covariant curvature formalism}

We employ the following metric form\footnote{In Ref \cite{Kanno} it is assumed that   
$\phi(y,x)$ does not depend on $y$. However, as seen later, the assumption can be 
removed.}
%
\begin{eqnarray}
ds^2=e^{2\phi(y,x)}dy^2+q_{\mu\nu}(y,x)dx^\mu dx^\nu.
\end{eqnarray}
%
In the above it is supposed that 
the positive and negative tension branes are located at $y=0$ and $y=y_0$, 
respectively. The proper distance between two branes is given by $d_0(x)=\int_0^{y_0} dy 
e^{\phi (y,x)}$. 
$q_{\mu\nu}(y,x)$ is the induced metric of $y=$constant hypersurfaces. 

We follow the geometrical procedure(covariant curvature formalism) developed 
in Ref. \cite{Tess,Roy}. For simplicity, we do not include the bulk fields except for 
the bulk consmological constant. 
From the Gauss-Codacci equations, first of all, we have two key equations
%
\begin{eqnarray}
{}^{(4)}G^\mu_\nu & = & \frac{3}{\ell^2} \delta^\mu_\nu+KK^\mu_\nu-K^\mu_\alpha K^\alpha_{\nu}
\nonumber \\
& & -\frac{1}{2}\delta^\mu_\nu (K^2-K^\alpha_\beta K^\beta_\alpha)-E^\mu_\nu \label{Gauss}
\end{eqnarray}
%
and
%
\begin{eqnarray}
D_\mu K^\mu_\nu-D_\mu K =0,
\end{eqnarray}
%
where $D_\mu$ is the covariant derivative with respect to $q_{\mu\nu}$. $\ell$ is the bulk  
curvature radius. 
${}^{(4)}G_{\mu\nu}$ is the 4-dimensional Einstein tensor with respect to 
$q_{\mu\nu}$, $K_{\mu\nu}$ is the extrinsic curvature of $y=$constant hypersurfaces 
defined by 
%
\begin{eqnarray}
K_{\mu\nu}=\frac{1}{2}\mbox \pounds_n q_{\mu\nu}=\nabla_\mu n_\nu+n_\mu D_\nu \phi,
\end{eqnarray}
%
where $n=e^{-\phi}\partial_y$. Here note that $a^\mu= n^\nu \nabla_\nu n^\mu=-D^\mu \phi (y,x)$.  
$E_{\mu\nu}$ is a part of the projected Weyl tensor defined by 
%
\begin{eqnarray}
E^\mu_\nu & = & {}^{(5)}C_{\mu\alpha\nu\beta}n^\alpha n^\beta \nonumber \\ 
& = &  -D^\mu D_\nu \phi - D^\mu \phi D_\nu \phi - \mbox \pounds_n K^\mu_\nu-K^\mu_\alpha K^\alpha_\nu
 \nonumber \\
& & +\frac{1}{\ell^2}\delta^\mu_\nu, \label{defE}
\end{eqnarray}
%
where ${}^{(5)}C_{\mu\nu\alpha\beta}$ is the 5-dimensional Weyl tensor. 

The junction conditions on the branes are 
%
\begin{eqnarray}
[K^\mu_\nu-\delta^\mu_\nu K ]_{y=0}=-\frac{\kappa^2}{2}\biggl(-\sigma_1 \delta^\mu_\nu+T^{\mu}_{1~\nu} \biggr) 
\end{eqnarray}
%
and
%
\begin{eqnarray}
[K^\mu_\nu-\delta^\mu_\nu K ]_{y=y_0}=\frac{\kappa^2}{2}\biggl(-\sigma_2 \delta^\mu_\nu+T^{\mu}_{2~\nu} \biggr). 
\end{eqnarray}
%
$T^\mu_{1~\nu}$ and $T^\mu_{2~\nu}$ are the energy-momentum tensor localised on the 
positive and negative branes. $\sigma_1$ and $\sigma_2$ are the brane tensions. 
If one substitutes the above conditions to Eq (\ref{Gauss}), 
we might be able to derive the Einstein equation on the brane. Indeed, this was successful 
procedure for the single brane\cite{Tess}. This is because $E_{\mu\nu}$ comes from 
just Kaluza-Klein modes and vanishes at the low energy\cite{Tama,Tess}. 
For the two brane systems, on the other hand, we have to carefully 
evaluate $E_{\mu\nu}$ due to the existence of the radion fields. 
Otherwise, we have a wrong prediction on the gravity on the 
branes. So we need the evolutional equation for $E_{\mu\nu}$ in the bulk. Even for the 
low energy scale, we learned from the linealised theory\cite{Tama} that $E_{\mu\nu}$ is not 
negligible. 

To evaluate $E_{\mu\nu}$ in the bulk, we derive its evolutional equation. The result is 
%
\begin{eqnarray}
\mbox \pounds_n E_{\alpha\beta}  & = & D^\mu B_{\mu (\alpha\beta)}+K^{\mu\nu}{}^{(4)}C_{\mu\alpha\nu\beta}
+4K^\mu_{(\alpha}E_{\beta) \mu} \nonumber \\ 
& & -\frac{3}{2}KE_{\alpha\beta}-\frac{1}{2}q_{\alpha\beta}K^{\mu\nu}E_{\mu\nu}
+2D^\mu \phi B_{\mu (\alpha\beta)}
 \nonumber \\
& & +2\tilde K^\mu_\alpha \tilde K_{\mu\nu} \tilde K^\nu_\beta-\frac{7}{6}\tilde K_{\mu\nu}
\tilde K^{\mu\nu}\tilde K_{\alpha\beta} \nonumber \\
& & -\frac{1}{2}q_{\alpha\beta}\tilde K_{\mu\nu}
\tilde K^\mu_\rho \tilde K^{\rho\nu}, \label{evoE}
\end{eqnarray}
%
where $B_{\mu\nu\alpha}=q_\mu^\rho q_\nu^\sigma {}^{(5)}C_{\rho\sigma\alpha\beta}n^\beta$ and 
$\tilde K_{\mu\nu}=K_{\mu\nu}-\frac{1}{4}q_{\mu\nu}K$. 
Since the right-hand side contains $B_{\mu\nu\alpha}$, $K_{\mu\nu}$ and 
${}^{(4)}C_{\mu\nu\alpha\beta}$, we also need their evolutional equations. After long calculation 
we are resulted in 
%
\begin{eqnarray}
& & \mbox \pounds_n {}^{(4)}R_{\mu\nu\alpha\beta}+2{}^{(4)}R_{\mu\nu\rho [\alpha}K^\rho_{\beta]}
+2D_{[\mu}B_{|\alpha\beta |\nu]} \nonumber \\
& &~~~ +2(D_\mu D_{[\alpha} \phi+ D_\mu \phi D_{[\alpha} \phi)K_{\beta]\nu} \nonumber \\
& & ~~~-2(D_{\nu}D_{[\alpha} \phi -D_\nu \phi D_{[\alpha}\phi )K_{\beta ] \mu} 
\nonumber \\
& &~~~ -2 B_{\alpha \beta [ \mu }D _{\nu ]} \phi -2 
B_{\mu \nu [ \alpha } D_{\beta ]} \phi =0
\end{eqnarray}
%
%
\begin{eqnarray}
& & \mbox \pounds_n  B_{\mu\nu\alpha}+2D_{[\mu} E_{\nu ]\alpha}
+2D_{[\mu}\phi E_{\nu]\alpha} \nonumber \\
& & ~~~- B_{\mu\nu\beta}K^\beta_\alpha+2 B_{\alpha\beta [\mu}
K_{\nu]}^\beta \nonumber \\
& &~~~ +({}^{(4)}R_{\mu\nu\alpha\beta}-K_{\mu\alpha}K_{\nu\beta}
+K_{\mu\beta}K_{\nu\alpha})D^\beta \phi =0
\end{eqnarray}
%
and
%
\begin{eqnarray}
e^{-\phi}\partial_y K^\mu_\nu & = & -D^\mu D_\nu \phi-D^\mu \phi D_\nu \phi  \nonumber \\
& & -K^\mu_\alpha 
K^\alpha_\nu+\frac{1}{\ell^2}\delta^\mu_\nu -E^\mu_\nu. \label{evoK}
\end{eqnarray}
%
Eq (\ref{evoK}) is just the rearrangement of Eq (\ref{defE}). The derivation 
is basically same with that in Ref \cite{Tess}.   
 
The junction condition directly implies the boundary condition on the branes for 
$K_{\mu\nu}$ and $B_{\mu\nu\alpha}$ because of 
%
\begin{eqnarray}
B_{\mu\nu\alpha}=2D_{[\mu} K_{\nu]\alpha}.
\end{eqnarray}
%

\section{Derivation of low energy effective theory}

It is now ready to 
derive the low energy theory for two brane systems. To do so, as stressed in the previous 
section, we must know $E_{\mu\nu}$ and solve the equation for $E_{\mu\nu}$ in the bulk. 
By low energy we 
mean that the typical scale of the curvature scale($L$) on the brane is much larger 
than the bulk curvature scale($\ell$), that is, $L \gg \ell$. The dimensionless parameter 
is $\epsilon=(\ell /L)^2$ which is tacitly entered into the equations below. We expand 
$K^\mu_\nu$ and $E^\mu_\nu$ as 
%
\begin{eqnarray}
K^\mu_\nu= {}^{(0)}K^\mu_\nu + {}^{(1)}K^\mu_\nu+\cdots
\end{eqnarray}
%
and
%
\begin{eqnarray}
E^\mu_\nu= {}^{(1)}E^\mu_\nu+\cdots.
\end{eqnarray}
%

\subsection{0-th order}

At the 0-th order, the evolutional equation which we have to solve is only one for 
$K_{\mu\nu}$; 
%
\begin{eqnarray}
e^{-\phi}\partial_y {}^{(0)}K^\mu_\nu= \frac{1}{\ell^2}\delta^\mu_\nu-{}^{(0)}K^\mu_\alpha
 {}^{(0)}K^\alpha_\nu.
\end{eqnarray}
%
And $K_{\mu\nu}$ satisfies the constraint
%
\begin{eqnarray}
D_\mu {}^{(0)}K^\mu_\nu-D_\nu {}^{(0)}K =0.
\end{eqnarray}
%
It is easy to see that the solution is 
%
\begin{eqnarray}
{}^{(0)}K^\mu_\nu=-\frac{1}{\ell}\delta^\mu_\nu.
\end{eqnarray}
%
From the definition
%
\begin{eqnarray}
\frac{1}{2}e^{-\phi}\partial_y{}^{(0)}q_{\mu\nu}=-\frac{1}{\ell}{}^{(0)}q_{\mu\nu},
\end{eqnarray}
%
the metric at the 0th order becomes 
%
\begin{eqnarray}
{}^{(0)}q_{\mu\nu}(y,x)=e^{-2d(y,x)/\ell}h_{\mu\nu}(x),
\end{eqnarray}
%
where $h_{\mu\nu}(x)$ is a tensor field depending on only the coordinate $x$ on the brane 
and $d(y,x)=\int^y_0 e^{\phi (y',x)}dy'$.  

At this order the junction condition is 
%
\begin{eqnarray}
[{}^{(0)}K^\mu_\nu-\delta^\mu_\nu {}^{(0)}K ]_{y=0}=\frac{\kappa^2}{2} \sigma_1 \delta^\mu_\nu
\end{eqnarray}
%
and
%
\begin{eqnarray}
[{}^{(0)}K^\mu_\nu-\delta^\mu_\nu {}^{(0)}K ]_{y=y_0}=-\frac{\kappa^2}{2} \sigma_2 \delta^\mu_\nu.
\end{eqnarray}
%
Thus the junction condition implies 
the relation between the bulk curvature radius $\ell$ and the brane tension 
$\sigma_{1,2}$ as 
%
\begin{eqnarray}
\frac{1}{\ell}=\frac{1}{6}\kappa^2 \sigma_1=-\frac{1}{6}\kappa^2 \sigma_2.
\end{eqnarray}
%
This is just the fine-tuning of Randall-Sundrum models\cite{RSI,RSII}. 

\subsection{1st order}

At the first order, the Riemann tensor does appear in the basic equations. So we can expect that 
the Einstein field equation will be able to described at this order. Indeed, the Gauss equation 
of Eq (\ref{Gauss}) becomes 
%
\begin{eqnarray}
{}^{(4)}G^\mu_\nu=-\frac{2}{\ell} ({}^{(1)}K^\mu_\nu-\delta^\mu_\nu {}^{(1)}K)-{}^{(1)}E^\mu_\nu. 
\end{eqnarray}
%
The first term in the right-hand side will be easily written in terms of the energy-momentum 
tensor on the branes using the junction condition at this order. So the unknown tensor is 
${}^{(1)}E^\mu_\nu$. 

The evolutional equation which we must solve are
%
\begin{eqnarray}
e^{-\phi} \partial_y {}^{(1)}E_{\mu\nu}=\frac{2}{\ell}{}^{(1)}E_{\mu\nu}. \label{evoE1st} 
\end{eqnarray}
%
and
%
\begin{eqnarray}
e^{-\phi}\partial_y {}^{(1)}K^\mu_\nu  & = &   -(D^\mu D_\nu \phi+D^\mu \phi D_\nu \phi) \nonumber \\
& &  +\frac{2}{\ell}{}^{(1)}K^\mu_\nu  -{}^{(1)}E^\mu_\nu. \label{evoK1st}
\end{eqnarray}
%
Eq (\ref{evoE1st}) is easily solved as 
%
\begin{eqnarray}
{}^{(1)}E_{\mu\nu}=e^{2d(y,x)/\ell} e_{\mu\nu}(x),
\end{eqnarray}
%
or 
%
\begin{eqnarray}
{}^{(1)}E^\mu_{\nu}=e^{4d(y,x) /\ell} \hat e^{\mu}_\nu (x), \label{solE}
\end{eqnarray}
%
where $\hat e^{\mu}_\nu (x)= h^{\mu\alpha}e_{\alpha \nu}(x)$. 

Substituting the expression Eq (\ref{solE}) into Eq (\ref{evoK1st}), we can obtain the 
solution for ${}^{(1)}K^\mu_\nu$ easily;
%
\begin{eqnarray}
& & {}^{(1)}K^\mu_\nu (y,x) \nonumber \\
& & ~~~= e^{2d/\ell}{}^{(1)}K^\mu_\nu (0,x)-\frac{\ell}{2}
(1-e^{-2d/\ell}){}^{(1)}E^\mu_\nu (y,x) \nonumber \\
& & ~~~~~~~-\biggl[D^\mu D_\nu d-\frac{1}{\ell} \biggl( D^\mu d D_\nu d -\frac{1}{2}\delta^\mu_\nu 
(Dd)^2 \biggr) \biggr]
\label{solK1st}
\end{eqnarray}
%
For the comparison, note that $e_{\mu\nu}$ cannot be determined by the junction condition 
in the RS2 model\cite{Kanno2,Tess}. To do so we need the boundary condition near the Cauchy horizon.

\subsection{Low energy effective theory for two brane systems}

At the 1st order the junction condition on the positive and negative tension branes becomes  
%
\begin{eqnarray}
{}^{(1)}K^\mu_\nu (0,x)-\delta^\mu_\nu {}^{(1)}K(0,x)=-\frac{\kappa^2}{2}T^\mu_{1~\nu}
\end{eqnarray}
%
and
%
\begin{eqnarray}
{}^{(1)}K^\mu_\nu (y_0,x)-\delta^\mu_\nu {}^{(1)}K(y_0,x)=\frac{\kappa^2}{2}T^\mu_{2~\nu}. 
\label{jun1st}
\end{eqnarray}
%
Using the above Eq (\ref{jun1st}), the Gauss equation on the negative tension brane becomes 
%
\begin{eqnarray}
{}^{(4)}G^\mu_\nu & = & -\frac{2}{\ell} \biggl( {}^{(1)}K^\mu_\nu (y_0,x) -\delta^\mu_\nu {}^{(1)}K(y_0,x) \biggr)
\nonumber \\
& & ~~~~~~-{}^{(1)}E^\mu_\nu (y_0,x) \nonumber \\
& = & -\frac{\kappa^2}{\ell}T^\mu_{2~\nu}-{}^{(1)}E^\mu_\nu(y_0,x).
\end{eqnarray}
%

Using the expression of Eq (\ref{solK1st}), the junction condition on the 
negative tension brane is written as 
%
\begin{eqnarray}
& & {}^{(1)}K^\mu_\nu(y_0,x)-\delta^\mu_\nu {}^{(1)}K(y_0,x) \nonumber \\
& & ~~=-\frac{\kappa^2}{2}e^{2d_0/\ell}T^\mu_{1~\nu}\nonumber \\
& & ~~~~-\biggl(D^\mu D_\nu d_0 -D^\mu D_\nu d_0  \biggr) \nonumber \\
& & ~~~~+\frac{1}{\ell}\biggl( D^\mu d_0 D_\nu d_0 
+\frac{1}{2}\delta^\mu_\nu (D d_0)^2 \biggr) \nonumber \\
& & ~~~~-\frac{\ell}{2}(1-e^{-2d_0/\ell}){}^{(1)}E^\mu_\nu (d_0,x) \nonumber \\
& & ~~=\frac{\kappa^2}{2}T^\mu_{2~\nu}.
\label{jun2}
\end{eqnarray}
%
In the right-hand side of the first line, we used the 
solution of Eq (\ref{solK1st}). The second line is just junction condition.  
From Eq (\ref{jun2}), then, the tensor $e_{\mu\nu}$(and then ${}^{(1)}E_{\mu\nu}$) 
is completely fixed as  
%
\begin{eqnarray}
& & \frac{\ell}{2}(1-e^{-2d_0/\ell}) e^{4d_0/\ell}\hat e^\mu_\nu (x)\nonumber \\
& &~~~ =  -\frac{\kappa^2}{2}(e^{2d_0/\ell}T^\mu_{1~\nu}+T^\mu_{2~\nu}) \nonumber \\
& &~~~~~ -\biggl( D^\mu D_\nu d_0 -\delta^\mu_\nu D^2 d_0 \biggr)\nonumber \\
& & ~~~~~+\frac{1}{\ell}\biggl( D^\mu d_0 D_\nu d_0 +\frac{1}{2} \delta^\mu_\nu 
(D d_0)^2 \biggr). \label{finalE}
\end{eqnarray}
%
The trace of Eq (\ref{finalE}) gives the equation for $d_0$ because $e^\mu_\nu$
is traceless. In general the radion field is massive. 
Substituting Eq (\ref{finalE}) into the Gauss equation at the first order, we 
can obtain the effective equation on the branes.  
On the negative tension brane, we have 
%
\begin{eqnarray}
{}^{(4)}G^\mu_\nu & = & \frac{\kappa^2}{\ell} \frac{1}{\Phi}T^\mu_{2~\nu}+\frac{\kappa^2}{\ell}
\frac{(1+\Phi)^2}{\Phi}T^\mu_{1~\nu} \nonumber \\
& & +\frac{1}{\Phi}(D^\mu D_\nu \Phi-\delta^\mu_\nu D^2\Phi)
\nonumber \\
& & +\frac{\omega(\Phi)}{\Phi^2}\biggl(D^\mu \Phi D_\nu \Phi-\frac{1}{2}\delta^\mu_\nu 
(D\Phi)^2 \biggr),
\end{eqnarray}
%
where $\Phi=e^{2d_0/\ell}-1$ and $\omega(\Phi)=-\frac{3}{2}\frac{\Phi}{1+\Phi}$. As should be so, 
this is exactly same result obtained by Kanno and Soda\cite{Kanno}. 
We also derive the effective equation on the positive tension brane easily;
%
\begin{eqnarray}
{}^{(4)}G^\mu_\nu & = & \frac{\kappa^2}{\ell} \frac{1}{\Psi}T^\mu_{1~\nu}+\frac{\kappa^2}{\ell}
\frac{(1-\Psi)^2}{\Psi} T^\mu_{2~\nu} \nonumber \\
& & +\frac{1}{\Psi}(\hat D^\mu \hat D_\nu \Psi-\delta^\mu_\nu \hat D^2\Psi)
\nonumber \\
& & +\frac{\omega(\Psi)}{\Psi^2}\biggl(\hat D^\mu \Psi \hat D_\nu \Psi-\frac{1}{2}\delta^\mu_\nu 
(\hat D\Psi)^2 \biggr),
\end{eqnarray}
%
where $\Psi=1-e^{-2d_0/\ell}$, $\omega(\Psi)=\frac{3}{2}\frac{\Psi}{1-\Psi}$ and $\hat D$ is the 
covariant derivative with respect to the induced metric $h_{\mu\nu}$ on the positive tension 
brane.

\section{Summary}

In this paper we derived the gravitational equation on the branes for the two brane systems 
at the low energy using the covariant curvature formalism\cite{Tess} and low energy expansion 
scheme\cite{Kanno,Kanno2}. The theory obtained here is presumably applicable to the 
cosmology and non-linear gravity at low energy scales.  
What we have done here is the evaluation of $E_{\mu\nu}$ at the low energy. In Ref \cite{Tess}, 
we thought that the anti-gravity appears on the negative tension brane supposing $E_{\mu\nu}$ 
is negligible. However, this is not correct and $E_{\mu\nu}$ is not negligible even at the 
low energy. 

Here we should comment on the difference between the study in Ref \cite{Kanno} and the present one. 
In Ref \cite{Kanno} $E_{\mu\nu}$ appears as the ``constant of integration". Thus it is 
difficult to proceed the discussion while keeping the physical meaning. This is just 
because of the metric based approach. On the other hand, $E_{\mu\nu}$ explicitly enters into the basic 
equations in the covariant curvature formalism and its physican meaning is manifest. Yet, 
the evolutional equation for $E_{\mu\nu}$ is simple at the low energy limit. 

For simplicity, we focused on the two brane systems without the radion stabilisation. If one 
is serious about the gauge hierarchy problem, we must assume that we are living on the 
negative tension brane. In this case, the gravity on the negative tension brane is scalar-tensor 
type and the scalar coupling is not permitted from the experimental point of view\cite{Tama}. 
So we should reconstruct our formalism in two brane systems with the radion stabilisation 
mechanism. This issue is left for future study.  

We are also interested in the higher order effects. We can expect that the effective theory with 
higher order corrections is higher-derivative type due to the non-local feature of the brane 
world\cite{Kanno,Kanno2,Mukohyama}.

\section*{Acknowledgments}

We would like to thank Jiro Soda, Sugumi Kanno, Daisuke Ida and Roy Maartens 
for fruitful discussions. TS's work is supported by Grant-in-Aid for Scientific
Research from Ministry of Education, Science, Sports and Culture of
Japan(No. 13135208, No.14740155 and No.14102004).


\end{document}